# TapirXLA: Embedding Fork-Join Parallelism into the XLA Compiler in TensorFlow Using Tapir

Tao B. Schardl
*MIT Computer Science and Artificial Intelligence Laboratory*
32 Vassar Street Cambridge, MA 02139
neboat@mit.edu

Siddharth Samsi
*MIT Lincoln Laboratory*
240 Wood Street Lexington, MA 02421
sid@ll.mit.edu

*Abstract*—This work introduces TapirXLA, a replacement for TensorFlow's XLA compiler that embeds recursive fork-join parallelism into XLA's low-level representation of code. Machine-learning applications rely on efficient parallel processing to achieve performance, and they employ a variety of technologies to improve performance, including compiler technology. But compilers in machine-learning frameworks lack a deep understanding of parallelism, causing them to lose performance by missing optimizations on parallel computation. This work studies how Tapir, a compiler intermediate representation (IR) that embeds parallelism into a mainstream compiler IR, can be incorporated into a compiler for machine learning to remedy this problem.

TapirXLA modifies the XLA compiler in TensorFlow to employ the Tapir/LLVM compiler to optimize low-level parallel computation. TapirXLA encodes the parallelism within high-level TensorFlow operations using Tapir's representation of fork-join parallelism. TapirXLA also exposes to the compiler implementations of linear-algebra library routines whose parallel operations are encoded using Tapir's representation. We compared the performance of TensorFlow using TapirXLA against TensorFlow using an unmodified XLA compiler. On four neural-network benchmarks, TapirXLA speeds up the parallel running time of the network by a geometric-mean multiplicative factor of 30% to 100%, across four CPU architectures.

## I. INTRODUCTION

Machine-learning (ML) frameworks, including Caffe [1], Flux [2], MXNet [3], PyTorch [4], TensorFlow [5], and Theano [6] have emerged as popular environments for developing ML applications. Because of the high demands of ML applications on computing resources, ML frameworks employ a variety of technologies to improve the performance of ML applications, including compiler technology for optimizing the computation within ML applications. In addition, ML applications typically exhibit substantial parallelism, which ML frameworks seek to exploit for performance using hardware accelerators, including GPUs and TPUs [7], and by invoking software libraries, such as Eigen [8], Intel's MKL-DNN [9], cuDNN [10], and cuBLAS [11], that implement linear-algebra routines that have been optimized to exploit parallel computing hardware. ML frameworks use all three technologies — compilers, high-performance software libraries, and hardware accelerators — to execute ML applications efficiently.

This material is based upon work supported by the Assistant Secretary of Defense for Research and Engineering under Air Force Contract No. (FA8721-05-C-0002 and/or FA8702-15-D-0001). Any opinions, findings and conclusions or recommendations expressed in this material are those of the author(s) and do not necessarily reflect the views of the Assistant Secretary of Defense for Research and Engineering.

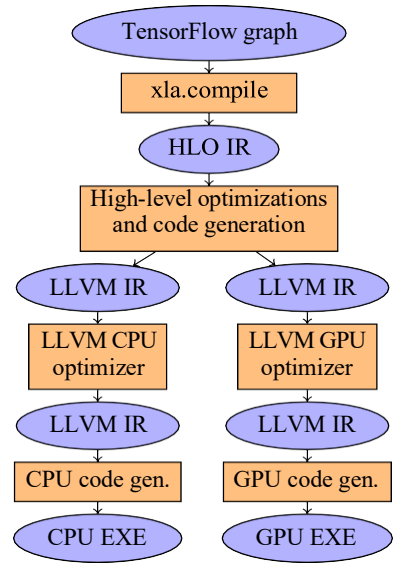

Fig. 1. Illustration of the high-level design of TensorFlow's XLA compiler. Ovals indicate the representation of the TensorFlow graph at different points in the compilation process. Rectangles denote stages of the compilation pipeline.

But compilers in ML frameworks struggle to effectively optimize low-level parallel computation within ML applications. For example, consider TensorFlow's XLA compiler [12], whose compilation pipeline for CPUs and GPUs is illustrated in Figure 1. As the figure shows, the XLA compiler compiles a TensorFlow graph — an ML network in TensorFlow — into executable machine code through a sequence of stages. The TensorFlow graph is first transformed into a high-level representation, called HLO IR, by a front-end, such as the `xla.compile` API [13]. Optimizations, such as operator fusion and common-subexpression elimination [14, Sec. 12.2], are performed on HLO IR before the graph is transformed into a lower-level representation, namely, LLVM's intermediate representation (IR) [15], for a target hardware architecture. LLVM performs additional optimizations on LLVM IR before it finally generates executable code from the optimized LLVM IR. Other ML compilers adopt a similar architecture, with multiple levels of IRs and a mainstream compiler, such as LLVM, operating on the lowest-level IR for final optimization and machine-code generation.

The LLVM optimizer in XLA fails to perform many ef-fective optimizations on parallel control flow in ML computations. (Other ML compilers suffer from similar problems.) The problem stems from the fact that parallel control flow is represented in LLVM IR as function calls into a parallel runtime system. Previous work [16] has observed that these function calls inhibit standard compiler optimizations, either because they are opaque — meaning the implementations of those functions are not exposed to the compiler — or they implement concurrency that compilers struggle to reason about. These missed optimizations can hurt the efficiency and parallel speedup of parallel code [16].

To mitigate this problem, as Figure 1 shows, XLA's high-level optimizer not only performs high-level optimizations on a TensorFlow graph, but it also decides how to implement these operations for a particular parallel back-end, such as a multicore CPU or a GPU. In particular, the high-level code-generator decides how to subdivide the computation into parallel tasks, in order to insert appropriate parallel runtime calls and low-level primitives to implement those parallel tasks for a given hardware back-end. These decisions are made before classic compiler optimizations in LLVM, such as loop unrolling or vectorization, have optimized the computation in each task. XLA uses heuristics to estimate how to subdivide the computation into parallel tasks, but subsequent compiler optimizations can upset these estimates, depending on how effectively different tasks are optimized.

A similar problem arises when ML compilers insert calls to libraries that implement optimized parallel linear-algebra kernels. These library calls inhibit compiler optimizations for similar reasons. When the library itself is opaque to the compiler, then the compiler cannot perform optimizations, such as function inlining [17, p. 536], constant propagation [17, p. 632], or common-subexpression elimination, on library routines based on the context in which they are called. When the source of a library routine is exposed to the compiler, the parallel library routine itself implements complex concurrency or calls to a parallel runtime system. Either way, these library calls inhibit compiler optimizations on the parallel computation in the application itself.

*The TapirXLA compiler*

This work studies the effect of enabling optimizations on parallel ML computations by embedding parallelism into an ML compiler and exposing parallel linear-algebra libraries to the compiler. Analysis and optimization of general parallel programs is a long-standing hard problem for compilers (see, for example, [18], [19], [20], [21], [22], [23], [24], [25], [26], [27], [28]). However, ML computations and linear-algebra routines often exhibit structured parallelism, specifically, recursive fork-join parallelism [29, Sec. 3.3], which includes loop parallelism. Previous work on Tapir [16] embeds recursive fork-join parallelism into the IR of a mainstream compiler to enable effective optimization on parallel computations. This work introduces TapirXLA, which integrates the Tapir/LLVM compiler [16] in place of the LLVM component in XLA to enable effective compiler optimizations on parallel ML computations.

Tapir supports a simple approach to compiling ML applications into efficient parallel code. Because the low-level Tapir/LLVM optimizer in TapirXLA optimizes parallel computation before inserting calls to a particular parallel runtime, the high-level optimizer need not use heuristics to decide how to partition the computation into parallel tasks. Instead, the high-level optimizer represents all logical fork-join parallelism in the ML computation using Tapir. The Tapir/LLVM optimizer then decides how to schedule and load-balance the computation among parallel tasks after it performs other optimizations on the parallel tasks.

TapirXLA exhibits substantial performance improvements compared to the XLA compiler in TensorFlow. Previous work [16] has shown that Tapir broadly improves the efficiency and parallel speedups of parallel programs, improving many benchmarks programs by factors of 5% to 25%. Although these are substantial performance gains for a compiler, TapirXLA demonstrates significantly larger performance improvements from applying Tapir to compile TensorFlow graphs. By embedding Tapir's representation of parallel computation into XLA and exposing the implementations of parallel linear-algebra kernels in Tapir, on four example networks, TapirXLA produces geometric-mean multiplicative speedups ranging from 30% to 100% across four multicore or manycore CPU architectures.

*Impact on machine-learning hardware*

Compiler technology, such as Tapir, for optimizing parallel computation has bearing on how ML frameworks utilize parallel hardware, including general-purpose multicore CPUs, GPUs, and specialized hardware accelerators such as TPUs [7]. Let us briefly survey the hardware currently used for machine learning.

Currently, the most widely used hardware platforms for training deep neural networks include NVIDIA GPUs. While open source ML frameworks such as TensorFlow and PyTorch tend to be optimized for the most commonly available hardware, the popularity of GPUs for neural network training has also been facilitated by the availability of highly optimized libraries such as cuBLAS and cuDNN that work on all GPUs ranging from laptops and desktops to high end systems such as the NVIDA DGX-2 that can achieve a peak performance of 2 PetaFLOPS.

While training neural networks in a High-Performance Computing center or in the cloud can leverage powerful GPUs, there are many applications where the availability of GPUs may be lacking due to power or form-factor constraints. For example, AI applications at the edge that run on IoT devices or on platforms with limited power capacity require specialized hardware for inference. GPUs also suffer from their limited memory capacity, compared to CPUs. The smaller amount of memory available on a GPU can lead to compromises such the use of small individual data sizes, down-sampling of training images, smaller than desired batch sizes, and changes to a model architecture. For example, a typical workflow may consist of extracting small image tiles from large, high resolution satellite or medical imaging data. Training a model then requires many data transfers between the CPU and the GPU, which is cost that can be minimized but not completely avoided. CPU-based systems can have significantly larger amounts of memory, from several hundred gigabytes to multiple terabytes, which can enable completely different kinds of applications. As a result, some networks, such as recurrent neural networks, have been observed to perform better on CPUs than GPUs [30].

| System | CPU | Cores |
|---|---|---|
| Summit | IBM Power9 | 2,397,824 |
| Sierra | IBM Power9 | 1,572,480 |
| Tianhe 2A | Intel Xeon E5 2692v2 | 10,649,600 |
| Piz Daint | Intel Xeon E5 2690v3 | 387,872 |
| Trinity | Intel Xeon E5 2698v3 and Xeon Phi 7250 | 979,072 |

Table I
Selected systems in the Top500 list [34] from November 2018.

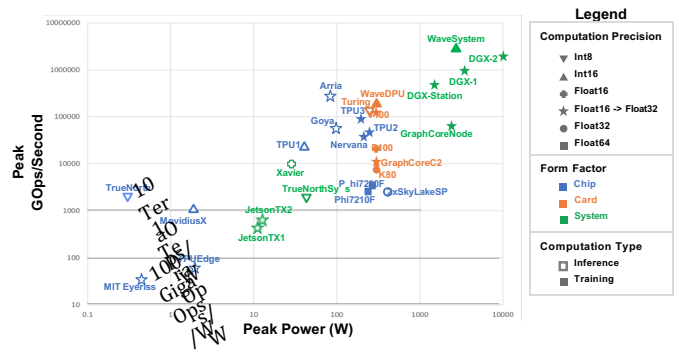

Fig. 2. Capabilities of current and planned processors for AI training and inference applications. The chart includes existing, commercially available hardware as well as processors being developed by academia. For more details, readers are referred to [36]. The variety of processors architectures available for AI applications makes it extremely challenging to optimize AI/ML frameworks for every possible processor architecture.

While GPUs remain the predominant hardware platform used for training deep neural networks, CPUs have also shown to be effective on machine-learning tasks. You et al. [31] demonstrated the training of ResNet50 and AlexNet deep convolutional networks on the ImageNet [32] dataset on 2,048 Intel Xeon Platinum 8160 processors. Shen et al. [33] demonstrated how CPUs can outperform GPUs on inference tasks. In addition, CPUs are cheaper, have broad software support and remain more readily available in every data center, cloud platform and in deployed systems. Every GPU system continues to have a host CPU that has the potential to be used for DNN training and inference. In recent years, high core count CPUs such as the Intel Xeon Phi and the Intel Xeon Scalable Processors have been integrated into many of the systems in the Top500 list. For example, some of the largest HPC systems in the world are built on a combination of thousands of CPU cores in conjunction with GPUs. Table I illustrates the availability of CPUs in large scale HPC systems.

There are many efforts underway in the academic [35] and commercial space to develop custom processors for training and inference applications [36]. Figure 2 shows an overview of the wide variety of hardware currently available as well as future hardware being developed in the academic and commercial world. The current trend is towards the development of custom hardware for dedicated applications. For example, the Google TPU v1 is an application specific integrated circuit (ASIC) designed primarily for inference applications [7]. Newer generations of the TPU are designed for both inference and training [37] and the Edge TPU is built for AI applications at the edge, where the inference applications are run at the point of data collection. Examples of such applications include autonomous vehicles and unmanned drones. Figure 2 graphs custom AI/ML architectures and capabilities, mapping peak performance vs. power usage.

This work on TapirXLA complements ongoing developments with parallel ML hardware. Embedding parallelism into an ML compiler affects how the compiler can optimize computation for different parallel hardware and target new hardware architectures. TapirXLA's performance results have bearing directly on the utility of CPUs for ML, specifically, in making CPUs more efficient and cost effective for ML applications. Moreover, Tapir's representation of parallel computation is not tied to a particular hardware architecture or parallel runtime system. Indeed, Tapir has been used for synthesizing efficient hardware for task-parallel programs [38]. By incorporating Tapir, an ML compiler can perform low-level optimizations on parallel computation that can benefit all hardware platforms and parallel runtime systems.

*Contributions*

This work explores the effectiveness of using Tapir to embed a parallelism into the compiler of an ML framework to enable low-level optimizations on ML applications. In particular, this paper makes the following contributions:

- We introduce TapirXLA, which incorporates Tapir into TensorFlow's XLA compiler. TapirXLA modifies XLA's high-level optimizer to encode fork-join parallelism in higher-level TensorFlow operations using Tapir.
- We developed parallel linear-algebra libraries whose parallel implementations are encoded using Tapir and exposed to TapirXLA.
- We evaluated TapirXLA on four neural networks and four multicore CPU systems. On these neural-network benchmarks, TapirXLA outperforms the XLA compiler in TensorFlow by a geometric mean multiplicative factor of 30% to 100%, across the different CPU architectures.

The remainder of the paper is organized as follows. Section II provides background on XLA and the model of recursive fork-join parallelism supported by Tapir. Section III describes how TapirXLA modifies the XLA compiler to incorporate Tapir and leverage its ability to optimize low-level parallel computation. Section IV presents our empirical evaluation of TensorFlow with the prototype TapirXLA compiler.

Section V discusses related work, and Section VI offers some concluding remarks.

## II. BACKGROUND ON XLA AND TAPIR

This section presents background on TensorFlow's XLA compiler [13] and the Tapir compiler intermediate representation (IR) [16]. We overview the design of TensorFlow's XLA compiler for CPUs and GPUs, which employs both a high-level IR and LLVM IR. We review Tapir/LLVM [16], which embeds fork-join parallelism into LLVM IR [15].

*TensorFlow's XLA compiler*

TensorFlow's XLA compiler [13] compiles TensorFlow graphs to run efficiently on different hardware architectures. Figure 1 illustrates the architecture of XLA, specifically, for targeting CPUs and GPUs. (XLA can also target TPUs [7] using a similar compilation pipeline.) As the figure shows, XLA optimizes TensorFlow graphs via transformations through multiple intermediate representations (IRs). XLA's high-level optimizer operates on HLO IR, which represents the TensorFlow graph as a data-flow graph of high-level operations on tensors. The high-level optimizer performs various optimizations on HLO IR, including operator fusion, before transforming the optimized HLO IR to LLVM IR.

When it generates LLVM IR, XLA exploits parallelism within the operations in the HLO IR. XLA generates LLVM IR to implement each operation in the HLO IR for the chosen hardware back-end, e.g., using appropriate calls into appropriate parallel CPU or GPU runtime libraries. To generate efficient parallel code, during this translation process, XLA uses heuristics to decide how to subdivide computation among parallel tasks for the target hardware. XLA also inserts calls to parallel libraries for standard linear-algebra routines, such as matrix multiplication or 2D convolution, that have been optimized for the target hardware. By default, when targeting CPUs, XLA inserts calls to the multithreaded Eigen linear-algebra library [8].

*Recursive fork-join parallelism and Tapir/LLVM*

In its implementation of HLO IR operations, XLA exploits (recursive) fork-join parallelism, which allows subroutines to be spawned recursively in parallel and iterations of a parallel loop to execute concurrently. The execution and synchronization of fork-join parallel tasks is managed "under the covers" by a runtime system, such as the Cilk Plus [39] or OpenMP runtime systems [40], [41]. Many computations can be parallelized efficiently using this simple model of parallelism. For some HLO IR operations, XLA parallelizes the operations directly using fork-join parallelism. For operations implemented in optimized linear-algebra libraries, XLA allows the library to use any model of parallelism to implement the operation. Although some multithreaded linear-algebra libraries do not use fork-join parallelism, the linear-algebra algorithms themselves can be parallelized efficiently using fork-join parallelism.

Tapir/LLVM [16] embeds fork-join parallelism into the intermediate representation of the LLVM compiler. Tapir adds three instructions — detach, reattach, and sync — to LLVM IR to express fork-join parallel control flow. These instructions allow the existing compiler's analyses and optimizations for serial programs to effectively optimize parallel programs with only minimal modifications. Tapir also enables new compiler optimizations specifically designed for parallel code. The Tapir/LLVM compiler has shown to be effective at optimizing recursive fork-join parallel programs to improve their efficiency and parallel speedup.

## III. DESIGN OF TAPIRXLA

This section describes how TapirXLA makes use of the Tapir/LLVM compiler in place of LLVM in the compilation pipeline, to enable optimizations on parallel machine-learning operations. We describe the simple strategy that TapirXLA employs for compiling and optimizing TensorFlow graphs. We describe the optimized parallel linear-algebra routines TapirXLA uses, whose parallel implementations are exposed to the compiler for optimization.

To leverage the Tapir/LLVM compiler's ability to optimize parallel computation, TapirXLA consists of two main changes to the XLA compiler in TensorFlow. First, TapirXLA compiles operations in HLO IR into parallel implementations that use Tapir's instructions for fork-join parallelism. Second, rather than emit opaque calls to the Eigen linear-algebra library, as XLA does by default, TapirXLA emits calls to Tapir-based parallel implementations of these linear-algebra routines and exposes those implementations to Tapir/LLVM. Let us examine these two changes more closely.

*TapirXLA's compilation strategy*

By leveraging Tapir's ability to optimize parallel code, TapirXLA supports a straightforward and effective strategy to optimize TensorFlow graphs. The high-level optimizer identifies operations in HLO IR that can be parallelized using fork-join parallelism. Typically, an operation in HLO IR can be implemented simply using nested parallel loops, e.g., over the dimensions of the operation's tensor input. For operations not handled by an optimized library, TapirXLA emits a parallel implementation of the operator using Tapir's constructs for fork-join parallelism. In particular, these parallelizable HLO IR operations are translated into Tapir's simple representation of a parallel loop [16].

Unlike XLA, TapirXLA does not attempt to optimize the parallel implementations of HLO IR operations, but instead emits a fork-join implementation of each operation that exposes all of its logical parallelism. TapirXLA thus relies on the Tapir/LLVM compiler to optimize these routines efficiently. Tapir/LLVM optimizes these operations using the full suite of LLVM optimization passes, which have been minimally modified to optimize parallel code [16]. Tapir/LLVM also performs optimizations that specifically target Tapir constructs, including parallel-loop strip-mining, loop spawning [16], and small-task serialization, which serializes the execution of a parallel task that performs too little work to overcome scheduling overheads. These Tapir/LLVM optimizations are not specialized to ML applications, but instead are generally applicable to fork-join parallel programs. Finally, Tapir/LLVM lowers Tapir's parallel constructs in the code to a parallel runtime system, meaning that it replaces Tapir instructions with appropriate calls to a parallel runtime library.

*Exposing parallel linear-algebra routines*

To uncover additional opportunities to optimize parallel code, TapirXLA incorporates the implementation of the parallel linear-algebra library. Hence, when TapirXLA inserts a call to this library, the Tapir/LLVM optimizer can subsequently optimize the library routine based on the context in which it is called. We ensured that the linear-algebra library routines were parallelized using Tapir to enable such optimization. We developed these parallel linear-algebra routines using Cilk [39] and compiled these routines using Tapir/LLVM [42] to an LLVM bitcode that uses Tapir instructions to encode the fork-join parallel control flow. When compiling a TensorFlow graph, TapirXLA incorporates this bitcode into the LLVM IR it produces for the TensorFlow graph. To facilitate optimization of these routines within the ML computation, these routines are optimized minimally when generating the bitcode file.

IV. EVALUATION

This section describes the evaluation of TapirXLA in comparison to the original XLA compiler in TensorFlow. We describe the implementation of TapirXLA and the experimental setup to perform a fair comparison TapirXLA against XLA. We evaluated TapirXLA on a variety of multicore and manycore CPUs on the MIT Supercloud system [43], a heterogenous supercomputing system consisting of compute nodes with a variety of multicore and manycore processors.

*Implementation of TapirXLA*

We implemented TapirXLA by modifying XLA in TensorFlow r1.13, the latest stable release of TensorFlow at the time of writing. We modified XLA to incorporate a version of Tapir/LLVM based on version 7.0 of the LLVM compiler [15]. We used the productivity tool suite integrated with Tapir/LLVM [42], including the Cilksan nondeterminism detector, to verify the correctness of the implementation. To create a scientific control to compare against this implementation of TapirXLA, we built a version TensorFlow r1.13 with an unmodified XLA compiler from source that incorporates the same version of LLVM and use the same configuration settings. For the different test machines, our builds of TensorFlow support the subset of, AVX, AVX2, and the fused-multiply-add (FMA) operations supported on the target CPU.

The TapirXLA implementation uses the Cilk Plus runtime [39] to execute parallel tasks. Although Tapir/LLVM contains prototype back-ends for other parallel runtime systems, the Tapir/LLVM back-end for the Cilk Plus runtime system is the most stable back-end at the time of writing. TapirXLA ensures that all parallel operations it compiles use the same back-end parallel runtime system, including parallel operations in linear-algebra library routines. As a result, TapirXLA mitigates performance issues arising from multiple parallel runtime systems competing for processor cores at the same time.

*Performance comparison of TapirXLA versus XLA*

We evaluated TapirXLA on training four benchmark neural networks written in TensorFlow: a small convolutional neural network (CNN), two LSTMs, and a recommendation network (NCF) [44]. The two LSTM benchmarks represent LSTM networks for isolated digit recognition (referred to as LSTM1 in Table 3) and continuous speech recognition (LSTM2), based on implementations described in [45]. The recommendation network was obtained from the suite of TensorFlow official models [46] and was evaluated using the MovieLens 1-million dataset [47]. All networks were compiled using the `xla.compile` API [13] to invoke TensorFlow's compiler.

We evaluated TapirXLA and XLA on all networks on a variety of multicore and manycore CPUs in the MIT Supercloud system [43], including an Intel Xeon Gold, an Intel Xeon E5, an Intel Xeon Phi, and an AMD Opteron. We followed the TensorFlow guidelines [48] and to set the threads used for intra- and inter-op thread counts equal to the number of processor cores and sockets on the system, respectively. We used `taskset` to pin worker threads in the parallel runtime systems onto single processor chips (sockets) on the target system. We evaluated other settings of these parameters to verify that these settings yield the best parallel performance for TensorFlow using either XLA or TapirXLA.

Figure 3 presents our performance results comparing TapirXLA and XLA on the benchmark ML networks and processor hardware. As the figure shows, TapirXLA consistently outperforms XLA across all networks and hardware systems. The magnitude of the performance improvement varies between networks and systems. TapirXLA yields a geometric mean multiplicative speedup of $2.1$ on newer Intel Xeon E5 and Intel Xeon Gold processors, whereas on the older Intel Xeon Phi and AMD Opteron systems, TapirXLA yields a geometric mean multiplicative speedup of $1.3$. TapirXLA appears to speed up different networks similarly, as the network that TapirXLA speeds up the most on a given processor differs between processors.

V. RELATED WORK

This section overviews related work on software technologies to optimize machine-learning applications. By and large, this software technology employs optimized libraries for machine-learning tasks as well as compiler technology to perform domain-specific optimizations on machine-learning applications. We discuss how this work on incorporating Tapir complements these other efforts and can, in principle, be used in conjunction with these technologies.

Software libraries, such as Intel's MKL-DNN [9], cuBLAS [10], and cuDNN [11], have been developed that encode highly optimized implementations of common ML operations for different hardware architectures. ML frameworks can use these highly optimized library routines by calling them directly from the language in which the ML application is written. Alternatively, compilers in ML frameworks can insert calls to these libraries when they compile and optimize an ML application. The later approach is compatible with this work to integrate Tapir into an ML compiler. This work also shows that substantial performance improvements can be obtained when a call to a library routine is not opaque, and instead, the parallel routine is encoded in Tapir and exposed to the compiler for optimization.

| Processor | Compiler | CNN (img/s) | LSTM1 (s) | LSTM2 (s) | NCF (s) |
|---|---|---|---|---|---|
| Intel Xeon Gold 6252/N, 2.3 GHz, 24 Cores | XLA | 1765.18 | 208.41 | 2567.73 | 248.42 |
| | TapirXLA | 3201.43 | 111.05 | 1182.14 | 148.64 |
| | *Ratio* | 1.81 | 1.88 | 2.17 | 1.67 |
| Intel Xeon E5-2683 v3, 2.00 GHz, 14 Cores | XLA | 423.15 | 443.19 | 4303.69 | 240.70 |
| | TapirXLA | 1792.36 | 185.83 | 1764.95 | 195.36 |
| | *Ratio* | 4.23 | 2.38 | 2.43 | 1.23 |
| Intel Xeon Phi 7210, 1.30 GHz, 64 Cores | XLA | 450.99 | 949.37 | 7907.04 | 793.70 |
| | TapirXLA | 649.65 | 635.51 | 4947.71 | 703.26 |
| | *Ratio* | 1.32 | 1.49 | 1.59 | 1.12 |
| AMD Opteron 6274, 2.20 GHz, 8 Cores | XLA | 496.17 | 1204.23 | 13,982.45 | 353.63 |
| | TapirXLA | 669.48 | 800.96 | 10,885.32 | 305.17 |
| | *Ratio* | 1.35 | 1.50 | 1.28 | 1.15 |

Fig. 3. Performance comparison of TensorFlow using vanilla XLA compiler versus TapirXLA. Each column lists TensorFlow's CPU performance on a given benchmark network when using either vanilla XLA or TapirXLA. The performance of the CNN benchmark is measured in overall images per second, while the performance of the other networks is measured in total running time (seconds). The rows labeled "Ratio" give the ratio of performance improvement that TapirXLA exhibits over XLA. For the CNN benchmark, this ratio equals the images-per-second performance of TapirXLA divided by the that of XLA. For all other benchmarks, this ratio equals the running time of XLA divided by that of TapirXLA. All performance values represent the average of 10 runs.

Many ML frameworks employ compiler technology to perform domain-specific optimizations on ML applications. TensorFlow's XLA compiler [12] performs operator fusion and common-subexpression elimination on high-level operations in a TensorFlow graph. PyTorch's Glow compiler [49] represents the ML computation through mutliple levels of IR to perform differentiation and a variety of optimizations, including domain-specific optimizations, memory optimizations, and quantization. DLVM [50] provides a compiler infrastructure to perform domain-specific optimizations on tensor computations, such as algebraic simplification and compute-kernel fusion. The TVM/NNVM compiler stack [51], [52] extends Halide [53] to perform loop optimizations on machine-learning networks. Flux and Zygote [54] employ the compiler technology in Julia to optimize ML applications and perform efficient reverse-mode automatic differentiation. All of these compilers employ LLVM late in their compilation pipelines to perform low-level optimizations and code generation. Hence, for any of these compilers, one can in principle apply Tapir/LLVM in place of LLVM to perform low-level optimizations on parallel computation, as this work does for the XLA compiler.

Intel's nGraph [55] provides a unified compiler stack to perform optimizations on ML applications written in a variety of frameworks and to target a variety of hardware back-ends. It remains an open research topic how compiler technology such as Tapir that enables compiler optimizations on low-level parallel operations can be applied within nGraph and what performance benefits Tapir may provide.

## VI. CONCLUSION

To conclude, this section discusses potential impact of Tapir on machine-learning frameworks.

In principle, Tapir can make it easy for ML compilers to target new parallel hardware architectures or runtime systems. For example, to target a new architecture or runtime system, XLA's high-level code-generation stage must emit efficient implementations of HLO IR operations for that architecture or runtime system. Tapir/LLVM shifts this burden to developing a new back-end for lowering Tapir's three instructions for parallel control flow [16]. Such a back-end allows any compiler that incorporates Tapir/LLVM to use the new hardware or runtime system. Tapir also allows scheduling and load-balancing decisions for parallel code to be made after optimizations have been performed on that code.

One compelling question considers how Tapir can optimize GPU code. ML frameworks often employ GPUs for training for neural networks, and many of the same issues explored in this paper concerning ML compilers and library calls pertain to GPUs as well as CPUs. Tapir seems to be a promising technology to overcome those challenges with GPU code, especially as Tapir's instructions are not tied to a particular hardware architecture or parallel runtime system. Developing a GPU back-end for Tapir remains an open research question.

## ACKNOWLEDGMENTS

The authors acknowledge the MIT Lincoln Laboratory Supercomputing Center for providing HPC resources that have contributed to the research results reported in this paper.